\documentclass{article}

\usepackage{arxiv}

\usepackage[utf8]{inputenc} 
\usepackage[T1]{fontenc}    
\usepackage{hyperref}       
\usepackage{url}            
\usepackage{booktabs}       
\usepackage{amsfonts}       
\usepackage{nicefrac}       
\usepackage{microtype}      
\usepackage{lipsum}		
\usepackage{graphicx}
\usepackage{natbib}
\usepackage{doi}

\title{Exploring the Educational Landscape of AI: Large Language Models' Approaches to Explaining Conservation of Momentum in Physics}

\date{} 					

\author{ 
        Keisuke Sato \\
	National Institute of Technology, Ibaraki College\\
        Hitachinaka, Ibaraki, Japan\\
	\texttt{skeisuke@ibaraki-ct.ac.jp} \\
}

\renewcommand{\headeright}{Technical Report}
\renewcommand{\shorttitle}{\textit{Exploring the Educational Landscape of AI} }

\hypersetup{
pdftitle={Exploring the Educational Landscape of AI},
pdfsubject={physics.ed-ph, cs.CY, cs.AI},
pdfauthor={Keisuke Sato},
pdfkeywords={Physics Education, Artificial Intelligence, Large Language Models, Conservation of Momentum, Educational Technology, AI-assisted Learning, STEM Education},
}

\begin{document}
\maketitle

\begin{abstract}
The integration of Large Language Models (LLMs) in education offers both opportunities and challenges, particularly in fields like physics that demand precise conceptual understanding. This study examines the capabilities of six state-of-the-art LLMs in explaining the law of conservation of momentum, a fundamental principle in physics. By analyzing responses to a consistent, simple prompt in Japanese, we assess the models' explanatory approaches, depth of understanding, and adaptability to different educational levels.Our comprehensive analysis, encompassing text characteristics, response similarity, and keyword usage, unveils significant diversity in explanatory styles across models. ChatGPT4.0 and Coral provided more comprehensive and technically detailed explanations, while Gemini models tended toward more intuitive approaches. Key findings include variations in the treatment of critical concepts such as net force, and differing emphases on mathematical rigor and real-world applications.The results indicate that different AI models may be more suitable for various educational contexts, ranging from introductory to advanced levels. ChatGPT4.0 and Coral demonstrated potential for advanced discussions, while Gemini models appeared more appropriate for introductory explanations. Importantly, the study underscores the necessity of educator guidance in effectively leveraging these AI tools, as models varied in their ability to convey nuanced aspects of physical principles.This research establishes a foundation for understanding the educational potential of LLMs in physics, providing insights for educators on integrating these tools into their teaching practices. It also highlights the need for further investigation into AI-assisted learning in STEM fields, paving the way for more sophisticated applications of AI in physics education.

\end{abstract}


\section{Introduction}
The rapid evolution of Large Language Models (LLMs) has ushered in new possibilities across various domains, with education at the forefront of this transformation. The potential of LLMs to revolutionize physics education, in particular, presents both promising opportunities and unique challenges \cite{gan2023large}. As these AI models grow increasingly sophisticated, understanding their capabilities and limitations in elucidating complex physical concepts becomes paramount for both educators and students.
Recent research has demonstrated LLMs' ability to provide detailed scientific explanations, respond to follow-up inquiries, and aid in problem-solving \cite{gan2023large,yeadon2023impact}. However, the application of these models in specific domains, such as physics education, remains largely unexplored. Polverini and Gregorcic \cite{polverini2024how} have recently provided a comprehensive overview of how understanding LLMs can inform their use in physics education, highlighting both the potential and limitations of these tools. This knowledge gap is especially significant given the abstract nature of many physics concepts and the field's reliance on precise, conceptually grounded explanations.
While Large Language Models (LLMs) offer promising opportunities for scientific research and education, their integration into scientific practices raises significant ethical and practical concerns. Birhane et al. \cite{birhane2023science} emphasize the need for careful consideration and responsible usage of LLMs in science to ensure that good scientific practices and trust in science are not compromised. This cautionary perspective underscores the importance of a nuanced approach to integrating LLMs in physics education.
The present study aims to address this gap by conducting a comprehensive analysis of how different state-of-the-art LLMs explain the concept of conservation of momentum. This fundamental principle in physics serves as an ideal test case due to its conceptual depth and central role in understanding various physical phenomena. By examining the responses of multiple LLMs, including ChatGPT (versions 3.5 and 4.0), Cohere's Command R+, and Google's Gemini (versions 1.0 Pro, 1.5 Flash, and 1.5 Pro), we aim to uncover the inherent capabilities and tendencies of each model within the context of physics education.

\section{Experimental Methods}
\label{sec:headings}

\subsection{AI Models}
We evaluated six state-of-the-art large language models, selected for their prominence and diverse approaches to language processing. The data was collected between June 29, 2024, and July 3, 2024. The models examined were:
\begin{enumerate}
    \item ChatGPT 3.5 turbo
    \item ChatGPT 4.0 turbo
    \item Coral (Command R+)
    \item Gemini 1.0 Pro
    \item Gemini 1.5 Flash
    \item Gemini 1.5 Pro
\end{enumerate}
These models represent a wide spectrum of LLMs, enabling us to examine varied approaches to explaining physical concepts. To ensure consistency across responses, we utilized the default temperature setting (0.7).
\subsection{Experimental Design}
We prompted each model with the following question in Japanese: "Explain the conservation law of momentum." (Please explain the law of conservation of momentum.) The choice of Japanese as the prompt language allowed us to explore how these models handle complex scientific concepts in a non-English context.

This simple, consistent prompt was deliberately employed to unveil the inherent tendencies and capabilities of each AI model, independent of specific prompt engineering techniques. To account for potential variations in responses and ensure statistical robustness, we repeated the prompt 50 times for each model.

\subsection{Data Collection and Analysis}
We collected and stored responses from each model in separate CSV files. Our analysis comprised three main components:

\subsubsection{Text Characteristics Analysis}
We examined each response for:
   - Character count
   - Sentence count
   - Average sentence length
   - Formality ratio

This analysis provided insights into the basic linguistic features of each model's responses, illuminating their general approach to explanation.

\subsubsection{Similarity Analysis}
o evaluate the consistency and diversity of responses, we calculated:
   - Cosine similarity
   - Jaccard similarity
   - Latent Semantic Analysis (LSA) similarity
   - Average Levenshtein distance

These measures enabled us to quantify the consistency of each model's explanations and the degree of variation in its responses.

\subsubsection{Keyword Analysis for Depth of Explanation}
We tracked the frequency of eight key physics terms to assess the depth and comprehensiveness of explanations:
   - Vector concept (including "vector", "direction", "orientation")
   - Momentum (specifically as "motion impetus")
   - Time (in the context of time invariance)
   - Net force
   - Derivation
   - Conservation of energy
   - Inelastic collision
   - Quantum mechanics

The presence and frequency of these keywords served as indicators of the depth of understanding and explanation provided by each model. Particular attention was paid to the use of "net force" in assessing explanation accuracy. A comprehensive explanation should clarify that momentum is conserved not only in the absence of external forces but also when external forces are present yet their net force (vector sum) is zero. This nuanced understanding is crucial for a thorough explanation of closed systems in the context of momentum conservation.

Furthermore, we analyzed the variation in keyword usage complexity and depth to evaluate the models' adaptability to different educational levels (introductory, intermediate, and advanced).

For each metric, we computed descriptive statistics (mean and standard deviation) across the 50 responses per model. For similarity measures, we calculated the average similarity between all pairs of responses within each model's dataset.

This multi-faceted approach facilitated a comprehensive evaluation and comparison of different AI models' performance in explaining the law of conservation of momentum. By progressing from basic text characteristics to similarity analysis and finally to in-depth keyword analysis, we aimed to construct a holistic understanding of each model's explanatory capabilities, with a primary focus on the depth and quality of explanations as indicated by the use of key physics concepts.

Our analytical method was designed not only to reveal the models' ability to explain the concept accurately but also to assess their potential suitability for various educational contexts and learner levels in physics education.

\section{Results}
\label{sec:Results}
\subsection{Text Characteristics Analysis}
Our analysis of text characteristics unveiled distinct patterns in how different AI models structure their explanations of the law of conservation of momentum. Table 1 presents a summary of key metrics for each model.
\begin{table}
	\caption{Summary of Text Characteristics Across AI Models}
	\centering
	\begin{tabular}{lcccc}
		\toprule
		Model & Character Count & Sentence Count & Avg. Sentence Length & Formality Ratio \\
		\midrule
		ChatGPT3.5 & 274.78 (±66.07) & 5.52 (±1.46) & 50.45 (±6.78) & 1.00 (±0.00) \\
		ChatGPT4.0 & 771.32 (±88.75) & 14.32 (±2.90) & 55.51 (±10.15) & 1.00 (±0.00) \\
		Coral & 714.06 (±110.92) & 18.26 (±4.44) & 40.28 (±6.30) & 1.00 (±0.01) \\
		Gemini-1.0-pro & 634.56 (±86.11) & 14.92 (±2.86) & 43.43 (±6.32) & 0.99 (±0.02) \\
		Gemini-1.5-flash & 773.60 (±76.80) & 18.16 (±3.06) & 43.21 (±4.70) & 1.00 (±0.01) \\
		Gemini-1.5-pro & 848.20 (±124.20) & 24.16 (±5.72) & 36.55 (±7.74) & 1.00 (±0.01) \\
		\bottomrule
	\end{tabular}
        \\Note: Values are presented as mean (±standard deviation).
	\label{tab:model_comparison}
\end{table}

\subsubsection{Response Length}
Gemini-1.5-pro generated the most extensive responses, averaging 848.20 ±124.20 characters, while ChatGPT3.5 produced the most concise, with an average of 274.78 ±66.07 characters. This substantial disparity in length indicates that Gemini-1.5-pro tends towards more comprehensive explanations, whereas ChatGPT3.5 favors brevity.

\subsubsection{Sentence Structure}
Gemini-1.5-pro exhibited the highest average sentence count (24.16 ±5.72), suggesting a preference for more segmented explanations. Conversely, ChatGPT3.5 employed the fewest sentences (5.52 ±1.46), indicating a more condensed explanatory style. Notably, ChatGPT4.0 demonstrated the longest average sentence length (55.51 ±10.15 characters), potentially signifying more intricate sentence structures.

\subsubsection{3.1.3 Explanation Complexity}
The average sentence length provides insight into the complexity of explanations. ChatGPT4.0's longer sentences (55.51 ±10.15 characters) suggest more intricate explanations, possibly incorporating multiple concepts within each sentence. In contrast, Gemini-1.5-pro's shorter average sentence length (36.55 ±7.74 characters) might indicate a preference for simpler, more digestible sentence structures, despite having longer overall responses.

\subsubsection{Formality}
All models maintained a high formality ratio (\( \geq 0.99 \)), appropriate for educational content. This consistency across models suggests a uniform capability to produce formal, academic-style explanations of scientific concepts.

\subsubsection{Response Consistency}
To compare the consistency of responses across models with varying average lengths, we calculated the Coefficient of Variation (CV) for character count. The CV, expressed as a percentage, represents the ratio of the standard deviation to the mean, providing a normalized measure of dispersion.

\begin{table}
	\caption{Coefficient of Variation for Character Count}
	\centering
	\begin{tabular}{lccc}
		\toprule
		Model & Mean Character Count & Standard Deviation & CV (\%) \\
		\midrule
		ChatGPT3.5 & 274.78 (±66.07) & 66.07 & 24.05\% \\
		ChatGPT4.0 & 771.32 (±88.75) & 88.75 & 11.51\% \\
		Coral & 714.06 (±110.92) & 110.92 & 15.53\% \\
		Gemini-1.0-pro & 634.56 (±86.11) & 86.11 & 13.57\% \\
		Gemini-1.5-flash & 773.60 (±76.80) & 76.80 & 9.93\% \\
		Gemini-1.5-pro & 848.20 (±124.20) & 124.20 & 14.64\% \\
		\bottomrule
	\end{tabular}
	\caption*{Note: Values are presented as mean (±standard deviation).}
	\label{tab:cv_character_count}
\end{table}

This analysis reveals that:

1. Gemini-1.5-flash demonstrates the highest consistency in response length, with a CV of 9.93\%.
2. ChatGPT4.0 also shows high consistency (CV = 11.51\%), despite producing much longer responses than ChatGPT3.5.
3. ChatGPT3.5, despite having the shortest average responses, shows the highest variability (CV = 24.05\%), indicating less consistency in response length.
4. Gemini-1.5-pro, which produces the longest responses on average, shows moderate consistency (CV = 14.64\%), suggesting a balance between detailed explanations and response stability.

These findings indicate that the more advanced models (except for ChatGPT3.5) generally provide more consistent response lengths, which could be beneficial for providing reliable and predictable educational content. The high variability in ChatGPT3.5's responses might suggest that it adapts its explanation length more dramatically based on the specific aspects of momentum conservation it chooses to focus on in each response.

\subsection{Similarity Analysis}
To evaluate the consistency and diversity of responses across different AI models, we conducted a comprehensive similarity analysis employing four distinct measures: Cosine similarity, Jaccard similarity, Latent Semantic Analysis (LSA) similarity, and average Levenshtein distance. These metrics provide valuable insights into the consistency of each model's explanations of the law of conservation of momentum and the degree of variation in its responses.
\begin{table}
	\caption{Similarity Metrics for Different Models}
	\centering
	\begin{tabular}{lcccc}
		\toprule
		Model & Cosine Similarity & Jaccard Similarity & LSA Similarity & Avg. Levenshtein Distance \\
		\midrule
		ChatGPT3.5       & 0.034 & 0.083 & 0.202 & 252.71 \\
		ChatGPT4.0       & 0.109 & 0.058 & 0.300 & 717.17 \\
		Coral            & 0.092 & 0.096 & 0.335 & 610.55 \\
		Gemini-1.0-pro   & 0.056 & 0.087 & 0.262 & 560.08 \\
		Gemini-1.5-flash & 0.099 & 0.152 & 0.339 & 594.17 \\
		Gemini-1.5-pro   & 0.061 & 0.082 & 0.277 & 798.64 \\
		\bottomrule
	\end{tabular}
	\label{tab:similarity_metrics}
\end{table}

\subsection{Lexical Similarity: Cosine and Jaccard Measures}

Cosine and Jaccard similarities quantify the textual overlap between responses, with higher values indicating greater similarity.

\begin{itemize}
    \item ChatGPT4.0 exhibited the highest Cosine similarity (0.108615), suggesting a more consistent use of specific terms and phrases across its responses.
    \item Gemini-1.5-flash demonstrated the highest Jaccard similarity (0.151965), indicating the most consistent vocabulary usage among all models.
    \item ChatGPT3.5 showed the lowest Cosine similarity (0.034002), implying more diverse responses in terms of specific wording and phrasing.
\end{itemize}

\subsubsection{Semantic Similarity: LSA Measure}

LSA similarity captures semantic consistency, considering the underlying meaning rather than exact word matches.

\begin{itemize}
    \item Gemini-1.5-flash and Coral exhibited the highest LSA similarities (0.338636 and 0.335099 respectively), indicating more consistent semantic content across their responses.
    \item ChatGPT3.5 demonstrated the lowest LSA similarity (0.201909), suggesting more diverse semantic content in its explanations.
\end{itemize}

\subsubsection{Structural Similarity: Levenshtein Distance}

Levenshtein distance measures the number of single-character edits required to change one response into another, with lower values indicating greater similarity.

\begin{itemize}
    \item ChatGPT3.5 had the lowest average Levenshtein distance (252.713469), consistent with its shorter average response length.
    \item Gemini-1.5-pro showed the highest average Levenshtein distance (798.636735), reflecting its longer and more varied responses.
\end{itemize}

\subsubsection{Implications of Similarity Analysis}

\begin{enumerate}
    \item \textbf{Consistency vs. Diversity}: While ChatGPT4.0 and Gemini-1.5-flash demonstrate higher consistency in their responses, ChatGPT3.5 and Gemini-1.5-pro exhibit more diversity. This suggests that more advanced models (except ChatGPT3.5) tend to provide more stable and consistent explanations.
    \item \textbf{Semantic Coherence}: The higher LSA similarities for Gemini-1.5-flash and Coral indicate that these models maintain more consistent semantic content across their responses, potentially offering more reliable explanations of core concepts.
    \item \textbf{Response Variability}: The high Levenshtein distance for Gemini-1.5-pro, combined with its moderate similarity scores, suggests that while it provides diverse explanations, it still maintains a consistent core message about conservation of momentum.
    \item \textbf{Balance of Consistency and Diversity}: ChatGPT4.0 appears to strike a balance between consistency (high Cosine similarity) and diversity (moderate Levenshtein distance), potentially offering reliable yet adaptable explanations.
\end{enumerate}

\subsection{Keyword Analysis for Depth of Explanation}
To assess the depth and comprehensiveness of explanations provided by each AI model, we analyzed the frequency of eight key physics terms in their responses. This analysis offers insights into how effectively each model incorporates crucial concepts when explaining the law of conservation of momentum.

\begin{table}
	\caption{Frequency of Key Physics Terms Across AI Models}
	\centering
	\begin{tabular}{lcccccc}
		\toprule
		Keyword & ChatGPT3.5 & ChatGPT4.0 & Coral & Gemini-1.0-pro & Gemini-1.5-flash & Gemini-1.5-pro \\
		\midrule
		Vector concept & 4 & 48 & 27 & 19 & 36 & 44 \\
		Momentum  & 0 & 0 & 0 & 0 & 2 & 9 \\
		Time invariance & 3 & 31 & 42 & 42 & 6 & 5 \\
		Net force & 4 & 14 & 6 & 0 & 2 & 0 \\
		Derivation & 0 & 4 & 16 & 4 & 9 & 3 \\
		Conservation of energy & 3 & 9 & 8 & 6 & 2 & 5 \\
		Inelastic collision & 0 & 15 & 34 & 8 & 0 & 1 \\
		Quantum mechanics & 2 & 7 & 21 & 3 & 1 & 4 \\
		\bottomrule
	\end{tabular}
	\label{tab:frequency_keywords}
\end{table}

\subsubsection{Vector Concept}

ChatGPT4.0 and Gemini-1.5-pro demonstrated the highest frequency of vector-related terms (48 and 44 times, respectively). This suggests a stronger emphasis on the vectorial nature of momentum, indicating a more comprehensive understanding of its directional properties.

\subsubsection{Momentum as "Motion Impetus"}

Interestingly, only Gemini models, particularly Gemini-1.5-pro (9 times), used the term "motion impetus" to describe momentum. This could indicate an attempt to make the concept more accessible to beginners by using intuitive language.

\subsubsection{Time Invariance}

Coral and Gemini-1.0-pro emphasized time invariance the most (42 times each), highlighting an important aspect of momentum conservation that some other models mentioned less frequently.

\subsubsection{Net Force}

ChatGPT4.0 mentioned net force most frequently (14 times), potentially indicating a more precise explanation of when momentum is conserved. This is particularly important as it demonstrates an understanding that momentum is conserved not only when no external forces are present, but also when external forces are present but their net force is zero.

\subsubsection{Implications of Keyword Analysis}

\begin{enumerate}
    \item \textbf{Depth of Explanation}: ChatGPT4.0 and Coral generally provided the most comprehensive explanations, incorporating a wide range of key concepts. This aligns with their higher text similarity scores, indicating consistent, in-depth explanations.
    \item \textbf{Conceptual Focus}: Different models showed varying emphasis on specific concepts. For example, Gemini models focused more on basic conceptual explanations (e.g., "motion impetus"), while Coral tended towards more advanced topics (e.g., inelastic collisions, quantum mechanics).
    \item \textbf{Precision in Explanations}: The frequency of "net force" mentions, particularly by ChatGPT4.0, indicates a more precise understanding of the conditions for momentum conservation. This suggests that some models are better at conveying the nuances of the concept.
    \item \textbf{Adaptability to Different Levels}: The variation in keyword usage suggests that different models might be more suitable for different educational levels. For instance, Gemini models might be more appropriate for introductory explanations, while Coral and ChatGPT4.0 could be better suited for advanced discussions.
\end{enumerate}

These findings provide crucial insights into the depth and quality of explanations offered by each AI model, revealing their potential suitability for different educational contexts and learner levels in physics education.

\section{Discussion}
\subsection{Diversity in AI Models' Explanatory Approaches}

Our study revealed significant heterogeneity in how different AI models approach the explanation of conservation of momentum. This diversity presents both opportunities and challenges for physics education.

The analysis uncovered distinct characteristics among the AI models. ChatGPT4.0 and Coral demonstrated a propensity for comprehensive, technically detailed explanations, incorporating advanced concepts such as vector representation and time invariance. In contrast, Gemini models, particularly Gemini-1.5-pro, exhibited a tendency towards more intuitive terminology (e.g., "motion impetus"), potentially enhancing accessibility for novice learners. ChatGPT3.5 consistently produced concise explanations, which may be particularly suitable for quick reviews or introductory overviews.

This diversity in explanatory approaches opens up new avenues for tailored physics education. Advanced students may benefit from the mathematically rigorous approaches of ChatGPT4.0 or Coral, while novice learners might find the intuitive explanations of Gemini models more accessible. The concise explanations of ChatGPT3.5 could serve as effective primers or quick reference tools.

However, this diversity also presents challenges. Ensuring consistency in core concept explanation across different AI models, developing guidelines for educators on when and how to utilize different AI models effectively, and addressing potential misconceptions that may arise from varying explanatory styles are important considerations for future implementation.

\subsection{Depth and Accuracy of Physical Concept Explanation}

Our analysis of keyword usage and explanation depth revealed nuanced patterns in how AI models handle the complexities of momentum conservation. A particularly noteworthy finding was the varied treatment of the concept of net force. ChatGPT4.0's frequent mention of net force (14 times) indicates a more sophisticated understanding of momentum conservation conditions. This nuanced approach goes beyond the simplified "no external forces" explanation often found in introductory texts. The inclusion of scenarios where external forces are present but their net force is zero represents a more accurate and comprehensive explanation of closed systems.

Our findings highlight a crucial trade-off in physics education between simplification for accessibility and maintaining physical accuracy. Models like ChatGPT3.5 offer more straightforward explanations, which may be easier for beginners to grasp. Conversely, more advanced models like ChatGPT4.0 and Coral provide deeper, more nuanced explanations that better reflect the complexities of physical phenomena. This trade-off suggests the need for a scaffolded approach in AI-assisted physics education, progressing from simplified to more complex explanations as students advance.

\subsection{Educational Implications}

The integration of AI models into physics education presents significant opportunities for enhancing learning experiences across various educational levels. Our findings suggest a potential framework for AI model selection based on educational level. At the introductory level, Gemini models and ChatGPT3.5, with their intuitive and concise explanations, may be most appropriate. For intermediate levels, ChatGPT4.0 and Coral, emphasizing mathematical rigor and advanced concepts, could provide valuable depth. At advanced levels, models like Coral, which make connections to topics such as quantum mechanics, could stimulate theoretical discussions and interdisciplinary thinking.

AI models can contribute to the development of problem-solving skills by demonstrating varied approaches to problem-solving, providing step-by-step explanations of derivations and mathematical processes, and offering examples of how to apply theoretical concepts to real-world situations. This aligns with the findings of Polverini and Gregorcic \cite{polverini2024how}, who illustrated how properly engineered prompts can enhance ChatGPT's output on conceptual introductory physics problems, potentially improving students' understanding and problem-solving skills. The diversity in AI explanations also provides an opportunity to enhance students' critical thinking skills by encouraging comparison and contrast of explanations from different AI models, prompting students to identify strengths, weaknesses, and potential biases in AI-generated content, and developing skills in evaluating and synthesizing information from multiple sources.

Several strategies for integrating AI into physics curricula emerge from our findings. These include using AI-generated explanations as supplementary material to traditional textbooks and lectures, incorporating AI models into interactive learning platforms for personalized learning experiences, and utilizing AI explanations in formative assessments to gauge student understanding and adapt instruction accordingly.

\subsection{Future Research Directions and Broader Implications}

This study provides a crucial foundation for understanding AI utilization in physics education. Building on these insights, future research could explore more advanced directions. For instance, the development of specialized prompt designs for physics education might play a vital role in maximizing the educational effectiveness of AI models. Additionally, comparative studies on long-term learning outcomes using different AI models or prompting techniques are essential for accurately assessing the impact of AI-assisted education. However, as highlighted by Birhane et al. \cite{birhane2023science}, the integration of LLMs in scientific and educational contexts necessitates careful ethical considerations to maintain the integrity of scientific practices and public trust in science.

Furthermore, investigating optimal strategies for combining AI assistance with traditional teaching methods represents another important research avenue. Exploring the potential of AI models to adapt to individual student learning styles and needs could offer more personalized educational experiences.

The implications of this research extend beyond physics education. It suggests possibilities for AI integration in other STEM fields dealing with complex, abstract concepts, and paves the way for developing AI-assisted educational tools that cater to diverse learning needs and levels. Moreover, this research may influence the shaping of educational policies regarding the use of AI in academic settings.

As LLM technology continues to advance, there is significant potential for developing more sophisticated AI-assisted learning tools in physics education. Recent work by Jiang and Jiang \cite{jiang2024beyond} with their Physics-STAR framework demonstrates a promising direction for future research. Their study shows how structured LLM interactions can enhance students' deep learning and precise understanding of physics concepts, particularly in addressing complex, information-based problems. This approach not only improves students' performance but also their efficiency in reasoning tasks, suggesting that well-designed LLM-powered tutoring systems could play a crucial role in the future of personalized physics education.

\section{Conclusion}

This study provides pioneering insights into the potential of AI models in physics education, specifically in explaining the concept of conservation of momentum. Our findings highlight the diverse capabilities of different AI models, suggesting their potential for tailored educational applications across various learning levels. While our research offers a solid foundation for understanding AI utilization in physics education, it also underscores the need for careful integration and ongoing research to maximize the benefits of AI in educational settings.

Limitations of this study include its focus on a single physics concept and the absence of testing in real classroom environments. These points suggest directions for future research. Expanding the analysis to cover a wider range of physics concepts, investigating long-term learning outcomes with AI-assisted instruction, and exploring methods for integrating AI explanations with traditional teaching approaches are important areas for future study.

As AI continues to evolve, its role in enhancing physics education presents both exciting opportunities and important challenges. This study serves as a springboard for future research and practical applications in the broader context of AI-assisted learning. It suggests the potential for revolutionary changes in how we approach education in the digital age, emphasizing the need for continued careful and proactive exploration in this field.

\section*{Acknowledgements}

This research utilized Claude 3.5 Sonnet, an AI assistant developed by Anthropic, as a research support tool for manuscript preparation and discussion organization. To maintain objectivity and avoid potential conflicts of interest, Claude was intentionally excluded from the experimental data collection process, which involved other AI models.

\bibliographystyle{unsrtnat}
\bibliography{references}  






\end{document}